# Evaluation of Big Data based CNN Models in Classification of Skin Lesions with Melanoma

Prasitthichai Naronglerdrit[1] and Iosif Mporas[2]

[1] Department of Computer Engineering, Faculty of Engineering at Sriracha,
Kasetsart University Sriracha Campus, Chonburi, Thailand
prasitthichai@eng.src.ku.ac.th

[2] School of Engineering and Computer Science, University of Hertfordshire,
Hatfield AL10 9AB, United Kingdom
i.mporas@herts.ac.uk

**Abstract.** This chapter presents a methodology for diagnosis of pigmented skin lesions using convolutional neural networks. The architecture is based on convolutional neural networks and it is evaluated using new CNN models as well as re-trained modification of pre-existing CNN models were used. The experimental results showed that CNN models pre-trained on big datasets for general purpose image classification when re-trained in order to identify skin lesion types offer more accurate results when compared to convolutional neural network models trained explicitly from the dermatoscopic images. The best performance was achieved by re-training a modified version of ResNet-50 convolutional neural network with accuracy equal to 93.89%. Analysis on skin lesion pathology type was also performed with classification accuracy for melanoma and basal cell carcinoma being equal to 79.13% and 82.88%, respectively.

## 1    Introduction

Over the last years skin cancer cases are becoming to a greater extent common with more than 5M people been diagnosed with it in the USA. Roughly three quarters of the skin cancer cases resulting in death have been caused from melanomas thus making it the most dangerous type of cancer of the skin [1]. The region of the spot on the skin which is affected is named lesion area and the lesions of the skin are the first clinical symptoms of diseases like melanoma, chickenpox and others [2]. As regards melanoma the growth rate is most of the cases slow enough to allow to be removed relatively easily and with low hazard and cost if the melanoma lesion has been detected in an initial stage. When the pathology of the skin is found in an early stage then the rate of patients to survive is more than 95% [3, 4], but on the other hand if the pathology of the skin is found at a late stage then treatment is more difficult with the survival rate being dramatically reduced to approximately 13% [5].

Currently visual examination is considered as the standard clinical procedure for detection and identification of skin lesions [6]. One of the most popular protocols for medical assessment of the lesions on the skin is the 'ABCD' protocol, according to which the dermatologist examines the asymmetry (A), the border (B), the colour (C) and the diameter (D) of the spot on the skin. As regards asymmetry it is examined whether segments of the skin lesion area diverge in shape or not. Regarding border, it is examined if the edges of the corresponding area of the skin are irregular or blurred and, in some cases, if they have notches. The colour might be uneven and with various colourizations of black, brown and pink. As regards the diameter of the skin lesion, most melanomas have diameter of at least 6 millimetres while any change of the size, the shape or the diameter being essential information the corresponding medical staff needs to be aware of [7]. Medical examination of the skin of patients is carried out by doctors, general practitioners or with expertise in dermatology, and typically it is an examination requiring a lot of time. Moreover, the diagnosis of lesions of skin is a very subjective procedure [8], as diagnosis can be imprecise or incorrect or could outcome to quite dissimilar diagnosis even if it has been done by dermatologists with lots of experience. In addition to subjectivity, diagnosis supported by AI-based computational tools can result in reduced diagnostic accuracy when diagnostic tools are utilized by doctors with no previous appropriate training or enough experience as shown in previous study [2].

Due to the limits and the difficulty of the clinical assessment of skin lesions as described above and also due to the number of cases related to skin diseases which are rising every year the development of accurate computer-aided tools for automated dermatoscopic image processing and for classification of skin lesions for the analysis of skin spots related to melanomas or different skin pathologies is necessary. Furthermore, the utilization of the cutting edge technological achievements in the areas of digital image processing, AI with emphasis in deep learning (machine learning) and big data applied for skin cancer detection can make it feasible to allow the diagnosis of skin lesions avoiding the need for body contact with the skin of the patient [2] or even perform the diagnosis from distance by sending to the dermatologist a photo of the corresponding skin area through the Internet. Moreover, the cost of making a diagnosis and treating of other than melanoma types of skin cancer is noteworthy as for instance only in 2010 it costed AU$ 511 million to Australia [9] and at the same time the overall cost to the healthcare system of the United States for the cases of melanoma is calculated at approximately $ 7 billion annually [10]. Due to the cost caused to national healthcare systems and aiming at retaining their sustainability, in countries like the United Kingdom and Australia there has been a lot of attention given in general practitioners to improve their diagnostic performance in order to be able to precisely identify and diagnose skin pathologies and cancers [11].

Dermatoscopy, also reported in the literature as dermoscopy or epiluminescence microscopy, is a skin imaging method which is non-invasive. Dermatoscopic images are magnified and illuminated images of an area of the skin which offer high resolution analysis of the lesions on the skin [8], nevertheless software-aided analysis of skin images has a number of issues making it a difficult task, like the usually not quite high contrast between the region of the spot and the normal skin region around it which results in automatic segmentation of the skin lesion areas with limited accuracy. Except this, another problem is that quite often the melanoma and the non-melanoma skin spots appear to be visually similar between them thus increasing making it difficult to distinguish a lesion corresponding to melanoma from a non-melanoma area of the skin. In addition, the variation of the characteristics of the skin from one patient to another, e.g. the skin colour and the presence of veins and natural hairs, result in melanoma cases with highly different colour and texture characteristics [3].

Different approaches for automatic image-based skin lesion classification have been published in the literature over the last decade. Rubegni et al. [12] used neural networks to perform automated classification of pigmented skin lesions, specifically melanoma vs nevi, and 13 image features representing the lesions were used parameterizing the geometry, the colourization, the textures and colour clusters of the skin lesions. In [4] Sheha et al. presented a classification architecture of malignant and benign skin lesions and evaluated the performance of classifiers like neural networks, support vector machines (SVMs) and k-nearest neighbours when using geometry based, chromatic and texture image features. In [10] Alcón et al. performed skin lesion binary classification with two skin lesion types, malignant and benign lesions, with the corresponding images being collected from standard digital cameras and evaluated decision trees using the AdaBoost metaclassifier and image features related to the symmetry, edges/boundaries, colour and texture of the skin spot area. Refianti et al. presented in [13] a binary classification methodology for melanoma skin cancer and non-melanoma skin cancer and the methodology relied on a convolutional neural network (CNN) structure. In [14] Prigent et al. proposed a skin hyper-pigmentation lesions classification methodology using multi-spectral analysis of skin images the classification was implemented utilizing support vector machines. In [11] Chang et al. introduced a software diagnostic algorithm to process malignant and benign lesions of skin which is extracting shape, texture and colour features of the lesion area and was tested using support vector machines as classifier on a dataset consisting of typical digital photos of skin. In [15] Kawahara et al. performed classification of skin images using convolutional neural networks with deep features extracted from the CNNs. Capdehourat et al. presented in [16] an approach for classification of pigmented skin lesions using support vector machines and decision trees evaluated on a set of dermatoscopic images. In [2] Sumithra et al. presented an evaluation of support vector machines and k-nearest neighbours classifiers in the task of skin image segmentation and lesions classification with parameterization of the lesion area using colour, texture and histogram-based image features. Lehman et al. in [17] proposed an approach which is fusing 3 convolutional neural networks, each of them using distinct image transformations. In [3] Li et al. presented an architecture for classification of skin lesion method which was utilizing parameters extracted from deep convolutional neural networks and were combined with colour, morphological and textural features of the skin lesion area.

Mahdiraji et al. in [6] calculated boundary features which were processed by neural networks to perform classification of skin lesions using sets of images acquired from conventional cameras. Mishra et al. [18] and Jafari et al. [19] presented segmentation approaches which can isolate the skin spot (i.e. the region of interest) from the remaining non-lesion part of the skin images with the employment of convolutional neural networks. In [20] M. ur Rehman et al. utilized convolutional neural networks to perform image segmentation followed by classification of the extracted image segments using artificial neural networks. Kaymak et al. in [21] presented a two-step deep learning algorithm for hierarchical classification of malignant pigmented skin lesions with the first step being classification of the skin images to melanocytic or non-melanocytic and the second stage being classification of the malignant types.

In the present paper we examine different models of convolutional neural networks for identification of skin lesion types captured from digital skin images using dermatoscopy. Except this we evaluate well known CNN models for image classification which have been pre-trained using big data and for the needs of skin lesion classification have been modified and re-trained. The remaining of this chapter is organized as follows. In Section 2 the block diagram and steps of the evaluated methodology for the automatic classification and diagnosis of lesions of skin is described. Section 3 presents the evaluation setup and in Section 4 the experimental evaluation results are presented. Finally, in Section 5 conclusions are given.

## 2      Classification and Diagnosis of Skin Lesions using CNNs

For the classification and diagnosis of skin lesions convolutional neural networks were utilized for the classification stage after using image processing algorithms for image pre-processing and segmentation of the dermatoscopic image. The dermatoscopic images are in a first step pre-processed, with the pre-processing step consisting of filtering of the image using a median filter and then removing of any detected hair. Following the pre-processing step image segmentation is performed to the skin images in order to detect and segment the area of the skin lesion, i.e. the region of interest (ROI) in the image. Once detecting the skin region of interest the corresponding bounding box is defined and then used to crop the skin lesion image segment which is subsequently converted to a predefined image size and sent to a convolutional neural network for identification of the corresponding skin lesion type. The block diagram of the dermatoscopic images-based skin lesion classification using convolutional neural networks as classifier is illustrated in Fig. 1.

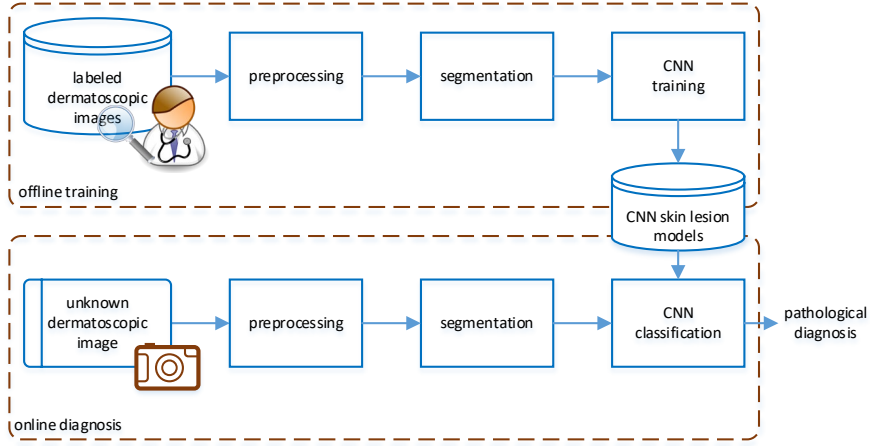

**Fig. 1.** Block diagram of the evaluated architecture for classification of skin lesions using convolutional neural networks.

During the offline training phase, a set of dermatoscopic images with known skin pathology labels, clinically verified by dermatologists, are pre-processed, segmented and the extracted ROIs used to train CNN models for classification of skin lesions. In the online diagnosis phase a new/unknown dermatoscopic image is processed following the same steps as in offline training and the CNN models assign a label to the corresponding skin lesion, thus performing online diagnosis. The pre-processing, segmentation and classification steps of the architecture are described in more detail below.

### 2.1 Dermatoscopic Image Preprocessing

As a first step we performed pre-processing of the dermatoscopic images for the purpose of hair removal [16, 22]. To automatically remove hair from the images hair detection is needed and subsequently is followed by inpainting of the images. In our implementation to detect the pixels of the dermatoscopic images that contained natural hair we utilized a median filter for smoothing and then applied bottom-hat filtering as in [23]. The bottom-hat (also referred to as black top-hat) filter is performing an image transformation, based on a morphological operation which is defined as

$$B_{hat}(f) = (f \bullet b) - f \tag{1}$$

here f is a grayscale image, b is a structuring element which was selected to be disk-shaped in the present evaluation. Both f and b are discrete, and $\bullet$ is the morphological operator of closing, defined as

$$f \bullet b = (f \oplus b) \ominus b \tag{2}$$

with $\oplus$ being the dilation operator and $\ominus$ being the erosion operator [24].

## 2.2    Dermatoscopic Image Segmentation

After pre-processing the skin images to detect and remove hair resizing to 256×256 pixels was applied and subsequently Gaussian filtering for smoothing of the pixel values of the images. As a next step, we used the active contour method as proposed by Chan-Vese [25] in order to segment the skin images to foreground and background. The active contour algorithm extracted the borders of the skin lesion and it was empirically found that 300 iterations was a fair trade-off between the achieved accuracy of segmentation vs the time needed for computations.

After the active contour segmentation, the detected skin lesions bounding box was found which had variable sizes. With the purpose of using the regions of interest, i.e. the segmented skin images, as input to a convolutional neural network all segmented images needed to be resized to a fixed size which was empirically selected equal to to 64×64 pixels. Except this, pixels' values were rescaled to the range [0, 1] for each of the 3 colours (RGB) to decrease big errors during subsequent processing from the activation function (ReLU in our implementation) of the convolutional neural network.

## 2.3    CNN-based Image Classification

The pre-processed, segmented and value-normalized skin images corresponding to the skin lesion areas (ROIs) were then sent as input to convolutional neural networks to classify them. The CNN classifier will assign a skin pathology label to each unknown dermatoscopic image so that automatic pathological diagnosis will be performed.

# 3    Experimental Setup

## 3.1    Data Description

To evaluate the skin lesions classification accuracy, the HAM10000 [26, 27] dataset was used. The HAM10000 database is a big dataset of digital images with skin lesions, which consist of 10015 skin images and all images have been labelled with their corresponding pathology type as shown in Table 1. The dermatoscopic images of the dataset is collected over a period of 20 years from two different sites, the Department of Dermatology at the Medical University of Vienna, Austria, and the skin cancer practice of Cliff Rosendahl in Queensland, Australia.

**Table 1.** Skin pathology types and number of instances in the HAM10000 dataset [26, 27].

| Skin Pathology | Number of images |
|---|---|
| Actinic keratosis (akiec) | 327 |
| Basal cell carcinoma (bcc) | 514 |
| Dermatofibroma (df) | 115 |
| Melanoma (mel) | 1113 |
| Nevus (nv) | 6705 |
| Pigmented benign keratosis (bkl) | 1099 |
| Vascular lesions (vasc) | 142 |
| Total | 10015 |

### 3.2    New CNN Models

As a baseline classification algorithm to classify the dermatoscopic images we used convolutional neural networks (CNNs) trained explicitly from the HAM10000 dataset images. The new CNN models were constructed after optimizing the convolutional layers number as presented in Table 2. As shown in the Table 2, the CNN models were built by using the pipeline of convolution layers and pooling layers. For the convolution layers, the filter size used was 3×3 and stride equal to 1. The pooling layers were used to reduce the size of the input for the next convolution layer and have filter size 2×2 and stride equal to 2.

In addition, the other hyper-parameters which were used in the setup of the CNN architecture consist of ReLU activation function, pooling layer implemented using max pooling, the optimizer using Adam and categorical cross entropy used as loss function. After the last pooling layer and dense layer, a dropout layer was used to avoid any over-fitting in the neural network. The classification of pigmented skin lesions was performed by a fully connected layer and softmax activation function.

**Table 2.** New CNN model architectures evaluated for skin lesion classification. The number of the model, i.e. 'Model-N', indicates the number of convolution layer and pooling layer sets in the CNN model.

| Layers | Output | Model 1 | Model 2 | Model 3 | Model 4 |
|---|---|---|---|---|---|
| Convolution | 64×64 | 3×3 conv, stride 1 | 3×3 conv, stride 1 | 3×3 conv, stride 1 | 3×3 conv, stride 1 |
| Pooling | 32×32 | 2×2 max pool, stride 2 | 2×2 max pool, stride 2 | 2×2 max pool, stride 2 | 2×2 max pool, stride 2 |
| Convolution | 32×32 | | 3×3 conv, stride 1 | 3×3 conv, stride 1 | 3×3 conv, stride 1 |
| Pooling | 16×16 | | 2×2 max pool, stride 2 | 2×2 max pool, stride 2 | 2×2 max pool, stride 2 |
| Convolution | 16×16 | | | 3×3 conv, stride 1 | 3×3 conv, stride 1 |
| Pooling | 8×8 | | | 2×2 max pool, stride 2 | 2×2 max pool, stride 2 |
| Convolution | 8×8 | | | | 3×3 conv, stride 1 |
| Pooling | 4×4 | | | | 2×2 max pool, stride 2 |
| Flatten | | 65536 | 16384 | 4096 | 1024 |
| Dropout | | | 0.5 dropout | | |
| Dense | 128 | | fully connected | | |
| Dropout | | | 0.5 dropout | | |
| Classification | | | 7 fully-connected, softmax | | |

The CNN models were constructed using Tensorflow and Keras packets, and the training process used the CUDA GPU computing for acceleration. The CNN models were trained for 50 epochs and batch normalization size equal to 16.

### 3.3    Pre-Trained CNN Models and their Modifications

Except training new CNN models we investigated the usage of pre-trained CNN models for image classification that available with the Keras packet, in particular the VGG [28], MobileNet [29], DenseNet [30], and ResNet [31] pre-trained CNN models. The pre-trained CNN models, the architectures of which were modified and used for pigmented skin lesions classification, were originally trained using ImageNet [32], which is a large images dataset consisting of 1,000 classes with 1.2 million training images and 50,000 validation images.

The VGG [28] network was constructed by the Visual Geometry Group, University of Oxford for the needs of ILSVRC-2014 competition. The VGG architecture is a sequential network which has 3×3 convolution layers with stride equal to 1 and 2x2 max-pooling layers with stride equal to 2. The original VGG network was trained by the ImageNet which has 1,000 classes with the input image size being equal to 224×224. In this paper we have used the VGG network with a depth of 16 layers (noted as VGG16) to perform as a classification model. In the modified VGG16 network, the last max-pooling layer was truncated, and the output was connected to the fully connected layer with softmax activation function for implementing the classification task as presented in Table 3.

**Table 3.** VGG16 [28] architecture and its modifications for skin lesion classification.

| Layers | Original VGG16 | | Modified VGG16 | |
|---|---|---|---|---|
| | output size | layer's structure | output size | layer's structure |
| Convolution | 224×224 | 3×3 conv, stride 1 | 64×64 | 3×3 conv, stride 1 |
| Convolution | 224×224 | 3×3 conv, stride 1 | 64×64 | 3×3 conv, stride 1 |
| Pooling | 112×112 | 2×2 max pool, stride 2 | 32×32 | 2×2 max pool, stride 2 |
| Convolution | 112×112 | 3×3 conv, stride 1 | 32×32 | 3×3 conv, stride 1 |
| Convolution | 112×112 | 3×3 conv, stride 1 | 32×32 | 3×3 conv, stride 1 |
| Pooling | 56×56 | 2×2 max pool, stride 2 | 16×16 | 2×2 max pool, stride 2 |
| Convolution | 56×56 | 3×3 conv, stride 1 | 16×16 | 3×3 conv, stride 1 |
| Convolution | 56×56 | 3×3 conv, stride 1 | 16×16 | 3×3 conv, stride 1 |
| Convolution | 56×56 | 3×3 conv, stride 1 | 16×16 | 3×3 conv, stride 1 |
| Pooling | 28×28 | 2×2 max pool, stride 2 | 8×8 | 2×2 max pool, stride 2 |
| Convolution | 28×28 | 3×3 conv, stride 1 | 8×8 | 3×3 conv, stride 1 |
| Convolution | 28×28 | 3×3 conv, stride 1 | 8×8 | 3×3 conv, stride 1 |
| Convolution | 28×28 | 3×3 conv, stride 1 | 8×8 | 3×3 conv, stride 1 |
| Pooling | 14×14 | 2×2 max pool, stride 2 | 4×4 | 2×2 max pool, stride 2 |
| Convolution | 14×14 | 3×3 conv, stride 1 | 4×4 | 3×3 conv, stride 1 |
| Convolution | 14×14 | 3×3 conv, stride 1 | 4×4 | 3×3 conv, stride 1 |
| Convolution | 14×14 | 3×3 conv, stride 1 | 4×4 | 3×3 conv, stride 1 |
| Pooling | 7×7 | 2×2 max pool, stride 2 | 2×2 | 2×2 max pool, stride 2 |
| Flatten | 25088 | | 2048 | |
| Dense | 4096 | | | |
| Dense | 4096 | | | |
| Classification | 1000 fully-connected, softmax | | 7 fully-connected, softmax | |

The MobileNet [29] is introduced by Google Inc. for mobile and embedded vision applications. The MobileNet uses the concept of depthwise separable convolutions in its light-weight architecture and to keep the channel of the network, the 1×1 filter is used as a pointwise convolution. The advantage of this architecture is the reduced number of computations needed, both during training the CNN model and during online testing. Table 4 shows the modified architecture of the MobileNet in which the input size has changed from 224×224 to 64×64. All convolutional layers are followed by batch normalization and ReLU for activation function.

**Table 4.** MobileNet [29] architecture and its modifications for skin lesion classification.

| Layers | Original MobileNet | | Modified MobileNet | |
|---|---|---|---|---|
| | output size | layer's structure | output size | layer's structure |
| Convolution | 112×112 | 3×3×32 conv, stride 2 | 32×32 | 3×3×32 conv, stride 2 |
| Convolution | 112×112 | 3×3 depthwise conv, stride 1 | 32×32 | 3×3 depthwise conv, stride 1 |
| Convolution | 112×112 | 1×1×64 conv, stride 1 | 32×32 | 1×1×64 conv, stride 1 |
| Convolution | 56×56 | 3×3 depthwise conv, stride 2 | 16×16 | 3×3 depthwise conv, stride 2 |
| Convolution | 56×56 | 1×1×128 conv, stride 1 | 16×16 | 1×1×128 conv, stride 1 |
| Convolution | 56×56 | 3×3 depthwise conv, stride 1 | 16×16 | 3×3 depthwise conv, stride 1 |
| Convolution | 56×56 | 1×1×128 conv, stride 1 | 16×16 | 1×1×128 conv, stride 1 |
| Convolution | 28×28 | 3×3 depthwise conv, stride 2 | 8×8 | 3×3 depthwise conv, stride 2 |
| Convolution | 28×28 | 1×1×256 conv, stride 1 | 8×8 | 1×1×256 conv, stride 1 |
| Convolution | 28×28 | 3×3 depthwise conv, stride 1 | 8×8 | 3×3 depthwise conv, stride 1 |
| Convolution | 28×28 | 1×1×256 conv, stride 1 | 8×8 | 1×1×256 conv, stride 1 |
| Convolution | 14×14 | 3×3 depthwise conv, stride 2 | 4×4 | 3×3 depthwise conv, stride 2 |
| Convolution | 14×14 | 1×1×512 conv, stride 1 | 4×4 | 1×1×512 conv, stride 1 |
| 5 x    Convolution | 14×14 | 3×3 depthwise conv, stride 1 | 4×4 | 3×3 depthwise conv, stride 1 |
|     Convolution | 14×14 | 1×1×512 conv, stride 1 | 4×4 | 1×1×512 conv, stride 1 |
| Convolution | 7×7 | 3×3 depthwise conv, stride 2 | 2×2 | 3×3 depthwise conv, stride 2 |
| Convolution | 7×7 | 1×1×1024 conv, stride 1 | 2×2 | 1×1×1024 conv, stride 1 |
| Convolution | 7×7 | 3×3 depthwise conv, stride 2 | 2×2 | 3×3 depthwise conv, stride 2 |
| Convolution | 7×7 | 1×1×1024 conv, stride 1 | 2×2 | 1×1×1024 conv, stride 1 |
| Classification | 1024 | 7×7 global average pool | 4096 | flatten |
| | 1000 fully-connected, softmax | | 7 fully-connected, softmax | |

The Dense Convolutional Network (DenseNet) [30] connects every layer to all next layers, while a conventional network connects only between previous layer to next layer. The block of the connections which connects each layer to other layers in feed-forward architecture noted as "dense block" as shown in Figure 2.

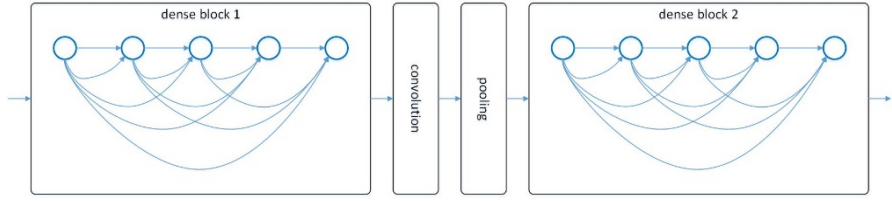

**Fig. 2.** DenseNet [30] with two dense blocks and transition layer between the dense blocks consisting of a convolution and pooling layer.

The DenseNet has a 3×3 filter size at the convolution layer, and between dense blocks a 1×1 convolution layer followed by a 2×2 average-pooling layer are used as transition layer. Finally, softmax classifier is attached at the end of the last convolution layer. DenseNet has four dense blocks, but different number of convolution layers inside each block. In this paper, we used the DenseNet-121, DenseNet-169, and DenseNet-201 CNNs where the number after DenseNet name is indicating the total number of convolution layers and fully-connected layers as shown in Tables 5, 6 and 7, respectively.

**Table 5.** DenseNet-121 [30] architecture and its modifications for skin lesion classification.

| Layers | Original DenseNet-121 | | Modified DenseNet-121 | |
|---|---|---|---|---|
| | output size | layer's structure | output size | layer's structure |
| Convolution | 112×112 | 7×7 conv, stride 2 | 32×32 | 7×7 conv, stride 2 |
| Pooling | 56×56 | 3×3 max pool, stride 2 | 16×16 | 3×3 max pool, stride 2 |
| Dense Block (1) | 56×56 | $\begin{bmatrix} 1\times1 \text{ conv} \\ 3\times3 \text{ conv} \end{bmatrix} \times 6$ | 16×16 | $\begin{bmatrix} 1\times1 \text{ conv} \\ 3\times3 \text{ conv} \end{bmatrix} \times 6$ |
| Transition (1) | 56×56 | 1×1 conv | 16×16 | 1×1 conv |
| Transition (1) | 28×28 | 2×2 average pool, stride 2 | 8×8 | 2×2 average pool, stride 2 |
| Dense Block (2) | 28×28 | $\begin{bmatrix} 1\times1 \text{ conv} \\ 3\times3 \text{ conv} \end{bmatrix} \times 12$ | 8×8 | $\begin{bmatrix} 1\times1 \text{ conv} \\ 3\times3 \text{ conv} \end{bmatrix} \times 12$ |
| Transition (2) | 28×28 | 1×1 conv | 8×8 | 1×1 conv |
| Transition (2) | 14×14 | 2×2 average pool, stride 2 | 4×4 | 2×2 average pool, stride 2 |
| Dense Block (3) | 14×14 | $\begin{bmatrix} 1\times1 \text{ conv} \\ 3\times3 \text{ conv} \end{bmatrix} \times 24$ | 4×4 | $\begin{bmatrix} 1\times1 \text{ conv} \\ 3\times3 \text{ conv} \end{bmatrix} \times 24$ |
| Transition (3) | 14×14 | 1×1 conv | 4×4 | 1×1 conv |
| Transition (3) | 7×7 | 2×2 average pool, stride 2 | 2×2 | 2×2 average pool, stride 2 |
| Dense Block (4) | 7×7 | $\begin{bmatrix} 1\times1 \text{ conv} \\ 3\times3 \text{ conv} \end{bmatrix} \times 16$ | 2×2 | $\begin{bmatrix} 1\times1 \text{ conv} \\ 3\times3 \text{ conv} \end{bmatrix} \times 16$ |
| Classification | 1024 | 7×7 global average pool | 4096 | flatten |
| | 1000 fully-connected, softmax | | 7 fully-connected, softmax | |

**Table 6.** DenseNet-169 [30] architecture and its modifications for skin lesion classification.

| Layers | Original DenseNet-169 | | Modified DenseNet-169 | |
|---|---|---|---|---|
| | output size | layer's structure | output size | layer's structure |
| Convolution | 112×112 | 7×7 conv, stride 2 | 32×32 | 7×7 conv, stride 2 |
| Pooling | 56×56 | 3×3 max pool, stride 2 | 16×16 | 3×3 max pool, stride 2 |
| Dense Block (1) | 56×56 | $\begin{bmatrix}1\times1\ \text{conv}\\3\times3\ \text{conv}\end{bmatrix}\times 6$ | 16×16 | $\begin{bmatrix}1\times1\ \text{conv}\\3\times3\ \text{conv}\end{bmatrix}\times 6$ |
| Transition (1) | 56×56 | 1×1 conv | 16×16 | 1×1 conv |
| Transition (1) | 28×28 | 2×2 average pool, stride 2 | 8×8 | 2×2 average pool, stride 2 |
| Dense Block (2) | 28×28 | $\begin{bmatrix}1\times1\ \text{conv}\\3\times3\ \text{conv}\end{bmatrix}\times 12$ | 8×8 | $\begin{bmatrix}1\times1\ \text{conv}\\3\times3\ \text{conv}\end{bmatrix}\times 12$ |
| Transition (2) | 28×28 | 1×1 conv | 8×8 | 1×1 conv |
| Transition (2) | 14×14 | 2×2 average pool, stride 2 | 4×4 | 2×2 average pool, stride 2 |
| Dense Block (3) | 14×14 | $\begin{bmatrix}1\times1\ \text{conv}\\3\times3\ \text{conv}\end{bmatrix}\times 32$ | 4×4 | $\begin{bmatrix}1\times1\ \text{conv}\\3\times3\ \text{conv}\end{bmatrix}\times 32$ |
| Transition (3) | 14×14 | 1×1 conv | 4×4 | 1×1 conv |
| Transition (3) | 7×7 | 2×2 average pool, stride 2 | 2×2 | 2×2 average pool, stride 2 |
| Dense Block (4) | 7×7 | $\begin{bmatrix}1\times1\ \text{conv}\\3\times3\ \text{conv}\end{bmatrix}\times 32$ | 2×2 | $\begin{bmatrix}1\times1\ \text{conv}\\3\times3\ \text{conv}\end{bmatrix}\times 32$ |
| Classification | 1664 | 7×7 global average pool | 6656 | flatten |
| | 1000 fully-connected, softmax | | 7 fully-connected, softmax | |

**Table 7.** DenseNet-201 [30] architecture and its modifications for skin lesion classification.

| Layers | Original DenseNet-201 | | Modified DenseNet-201 | |
|---|---|---|---|---|
| | output size | layer's structure | output size | layer's structure |
| Convolution | 112×112 | 7×7 conv, stride 2 | 32×32 | 7×7 conv, stride 2 |
| Pooling | 56×56 | 3×3 max pool, stride 2 | 16×16 | 3×3 max pool, stride 2 |
| Dense Block (1) | 56×56 | $\begin{bmatrix}1\times1\ \text{conv}\\3\times3\ \text{conv}\end{bmatrix}\times 6$ | 16×16 | $\begin{bmatrix}1\times1\ \text{conv}\\3\times3\ \text{conv}\end{bmatrix}\times 6$ |
| Transition (1) | 56×56 | 1×1 conv | 16×16 | 1×1 conv |
| Transition (1) | 28×28 | 2×2 average pool, stride 2 | 8×8 | 2×2 average pool, stride 2 |
| Dense Block (2) | 28×28 | $\begin{bmatrix}1\times1\ \text{conv}\\3\times3\ \text{conv}\end{bmatrix}\times 12$ | 8×8 | $\begin{bmatrix}1\times1\ \text{conv}\\3\times3\ \text{conv}\end{bmatrix}\times 12$ |
| Transition (2) | 28×28 | 1×1 conv | 8×8 | 1×1 conv |
| Transition (2) | 14×14 | 2×2 average pool, stride 2 | 4×4 | 2×2 average pool, stride 2 |
| Dense Block (3) | 14×14 | $\begin{bmatrix}1\times1\ \text{conv}\\3\times3\ \text{conv}\end{bmatrix}\times 48$ | 4×4 | $\begin{bmatrix}1\times1\ \text{conv}\\3\times3\ \text{conv}\end{bmatrix}\times 48$ |
| Transition (3) | 14×14 | 1×1 conv | 4×4 | 1×1 conv |
| Transition (3) | 7×7 | 2×2 average pool, stride 2 | 2×2 | 2×2 average pool, stride 2 |
| Dense Block (4) | 7×7 | $\begin{bmatrix}1\times1\ \text{conv}\\3\times3\ \text{conv}\end{bmatrix}\times 32$ | 2×2 | $\begin{bmatrix}1\times1\ \text{conv}\\3\times3\ \text{conv}\end{bmatrix}\times 32$ |
| Classification | 1920 | 7×7 global average pool | 7680 | flatten |
| | 1000 fully-connected, softmax | | 7 fully-connected, softmax | |

ResNet [31] was introduced by Microsoft Research team based on the residual learning framework. In residual networks shortcut connections between stacks of convolution layers are inserted, as illustrated in Figure 3. The ResNet has different number of convolution layers as shown in Tables 8, 9 and 10, respectively. For this evaluation we consider the ResNet-50, ResNet-101, and ResNet-152.

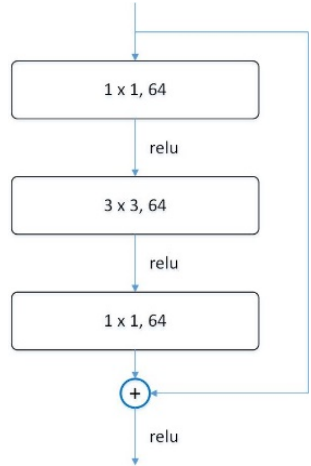

**Fig. 3.** Shortcut connection between a stack of 3 convolution layers of a residual network.

**Table 8.** ResNet-50 [31] architectures and their modifications for skin lesion classification.

| Layers | Original ResNet-50 | | Modified ResNet-50 | |
|---|---|---|---|---|
| | output size | layer's structure | output size | layer's structure |
| Convolution | 112×112 | 7×7 conv, stride 2 | 32×32 | 7×7 conv, stride 2 |
| Pooling | 56×56 | 3×3 max pool, stride 2 | 16×16 | 3×3 max pool, stride 2 |
| Convolution | 56×56 | $\begin{bmatrix} 1×1, 64 \\ 3×3, 64 \\ 1×1, 256 \end{bmatrix} × 3$ | 16×16 | $\begin{bmatrix} 1×1, 64 \\ 3×3, 64 \\ 1×1, 256 \end{bmatrix} × 3$ |
| Convolution | 28×28 | $\begin{bmatrix} 1×1, 128 \\ 3×3, 128 \\ 1×1, 512 \end{bmatrix} × 4$ | 8×8 | $\begin{bmatrix} 1×1, 128 \\ 3×3, 128 \\ 1×1, 512 \end{bmatrix} × 4$ |
| Convolution | 14×14 | $\begin{bmatrix} 1×1, 256 \\ 3×3, 256 \\ 1×1, 1024 \end{bmatrix} × 6$ | 4×4 | $\begin{bmatrix} 1×1, 256 \\ 3×3, 256 \\ 1×1, 1024 \end{bmatrix} × 6$ |
| Convolution | 7×7 | $\begin{bmatrix} 1×1, 512 \\ 3×3, 215 \\ 1×1, 2048 \end{bmatrix} × 3$ | 2×2 | $\begin{bmatrix} 1×1, 512 \\ 3×3, 215 \\ 1×1, 2048 \end{bmatrix} × 3$ |
| Classification | 2048 | 7×7 global average pool | 8192 | flatten |
| | 1000 fully-connected, softmax | | 7 fully-connected, softmax | |

**Table 9.** ResNet-101 [31] architectures and their modifications for skin lesion classification.

| Layers | Original ResNet-101 | | Modified ResNet-101 | |
|---|---|---|---|---|
| | output size | layer's structure | output size | layer's structure |
| Convolution | 112×112 | 7×7 conv, stride 2 | 32×32 | 7×7 conv, stride 2 |
| Pooling | 56×56 | 3×3 max pool, stride 2 | 16×16 | 3×3 max pool, stride 2 |
| Convolution | 56×56 | $\begin{bmatrix} 1\times1, 64 \\ 3\times3, 64 \\ 1\times1, 256 \end{bmatrix} \times 3$ | 16×16 | $\begin{bmatrix} 1\times1, 64 \\ 3\times3, 64 \\ 1\times1, 256 \end{bmatrix} \times 3$ |
| Convolution | 28×28 | $\begin{bmatrix} 1\times1, 128 \\ 3\times3, 128 \\ 1\times1, 512 \end{bmatrix} \times 4$ | 8×8 | $\begin{bmatrix} 1\times1, 128 \\ 3\times3, 128 \\ 1\times1, 512 \end{bmatrix} \times 4$ |
| Convolution | 14×14 | $\begin{bmatrix} 1\times1, 256 \\ 3\times3, 256 \\ 1\times1, 1024 \end{bmatrix} \times 23$ | 4×4 | $\begin{bmatrix} 1\times1, 256 \\ 3\times3, 256 \\ 1\times1, 1024 \end{bmatrix} \times 23$ |
| Convolution | 7×7 | $\begin{bmatrix} 1\times1, 512 \\ 3\times3, 215 \\ 1\times1, 2048 \end{bmatrix} \times 3$ | 2×2 | $\begin{bmatrix} 1\times1, 512 \\ 3\times3, 215 \\ 1\times1, 2048 \end{bmatrix} \times 3$ |
| Classification | 2048 | 7×7 global average pool | 8192 | flatten |
| | 1000 fully-connected, soft-max | | 7 fully-connected, softmax | |

**Table 10.** ResNet-152 [31] architectures and their modifications for skin lesion classification.

| Layers | Original ResNet-152 | | Modified ResNet-152 | |
|---|---|---|---|---|
| | output size | layer's structure | output size | layer's structure |
| Convolution | 112×112 | 7×7 conv, stride 2 | 32×32 | 7×7 conv, stride 2 |
| Pooling | 56×56 | 3×3 max pool, stride 2 | 16×16 | 3×3 max pool, stride 2 |
| Convolution | 56×56 | $\begin{bmatrix} 1\times1, 64 \\ 3\times3, 64 \\ 1\times1, 256 \end{bmatrix} \times 3$ | 16×16 | $\begin{bmatrix} 1\times1, 64 \\ 3\times3, 64 \\ 1\times1, 256 \end{bmatrix} \times 3$ |
| Convolution | 28×28 | $\begin{bmatrix} 1\times1, 128 \\ 3\times3, 128 \\ 1\times1, 512 \end{bmatrix} \times 8$ | 8×8 | $\begin{bmatrix} 1\times1, 128 \\ 3\times3, 128 \\ 1\times1, 512 \end{bmatrix} \times 8$ |
| Convolution | 14×14 | $\begin{bmatrix} 1\times1, 256 \\ 3\times3, 256 \\ 1\times1, 1024 \end{bmatrix} \times 36$ | 4×4 | $\begin{bmatrix} 1\times1, 256 \\ 3\times3, 256 \\ 1\times1, 1024 \end{bmatrix} \times 36$ |
| Convolution | 7×7 | $\begin{bmatrix} 1\times1, 512 \\ 3\times3, 215 \\ 1\times1, 2048 \end{bmatrix} \times 3$ | 2×2 | $\begin{bmatrix} 1\times1, 512 \\ 3\times3, 215 \\ 1\times1, 2048 \end{bmatrix} \times 3$ |
| Classification | 2048 | 7×7 global average pool | 8192 | flatten |
| | 1000 fully-connected, soft-max | | 7 fully-connected, softmax | |

All pre-trained CNN models of Keras packet were originally trained using ImageNet dataset, and modified for the needs of the present evaluation by changing the input size from 224×224 to 64×64 and connecting the last convolution layer with a flatten layer followed by a softmax fully-connected layer. Afterwards, the modified pre-trained models were re-trained using the HAM10000 dataset for skin lesions classification. The re-training process of the pre-trained CNN models was done using 50 epochs and batch normalization size equal to 16 by using Adam optimizer and categorical cross entropy as loss function.

## 4    Evaluation Results

The architecture for classification of skin lesions using convolutional neural networks presented in Section 2 was evaluated according to the experimental protocol described in Section 3. The performance of the evaluated CNN models was measured in terms of classification accuracy, i.e.

$$Accuracy = \frac{TP + TN}{TP + TN + FP + FN} \tag{3}$$

where $TP$ is the number of true positives, $TN$ is the number of true negatives, $FP$ is the number of false positives and $FN$ is the number of false negatives of the classified dermatoscopic images. To avoid overlap between the training and testing subsets of evaluated data a cross validation evaluation setup using 10 folds was employed.

The evaluation results of identification of skin lesion types from dermatoscopic images using new CNN models and re-trained models of modifications of pre-existing CNN models initially trained from big data are presented in Table 11. The best classification accuracy of the new CNN models and the best classification accuracy of the modified pre-existing CNN models are shown in bold.

**Table 11.** Dermatoscopic image classification accuracy (in percentages) for different CNN model architectures.

|            | CNN Model               | Accuracy (%) |
|------------|-------------------------|--------------|
| New        | CNN Model 1             | 72.68        |
| New        | CNN Model 2             | 76.38        |
| New        | CNN Model 3             | **76.83**    |
| New        | CNN Model 4             | 75.29        |
| Re-trained | VGG16 (modified)        | 85.19        |
| Re-trained | MobileNet (modified)    | 91.32        |
| Re-trained | DenseNet-121 (modified) | 91.21        |
| Re-trained | DenseNet-169 (modified) | 91.30        |
| Re-trained | DenseNet-201 (modified) | 89.56        |
| Re-trained | ResNet-50 (modified)    | **93.89**    |
| Re-trained | ResNet-101 (modified)   | 90.93        |
| Re-trained | ResNet-152 (modified)   | 93.52        |

As shown in Table 11 the top performing CNN model trained only with HAM10000 data ('New' CNN models) was CNN model 3 with classification accuracy equal to 76.83%. Among the evaluated pre-existing and modified CNN models the best classification accuracy was observed by the ResNet-50 with the accuracy equal to 93.89%, followed by the ResNet-152, the MobileNet, the DenseNet-169 and the DenseNet-121 CNN models with classification accuracies equal to 93.52%, 91.32%, 91.30% and 91.21%, respectively. The remaining big data-based CNN models achieved classification accuracy lower than 90%. All CNN models pretrained from big datasets for general purpose image classification outperformed the 4 evaluated CNN models trained explicitly from the evaluation dataset for skin lesion classification, thus indicating that the availability of large data collections is essential in training robust skin lesion classification models as well as that in the task of skin lesion classification CNN transferability is possible, i.e. pre-trained convolutional neural networks can successfully be modified and re-trained on smaller datasets of dematoscopic images and offer competitive classification accuracy.

In a second step we investigated the classification accuracy on skin pathology category level for the best performing CNN classification model, i.e. ResNet-50. The confusion matrix of the skin lesion classification using ResNet-50 is shown in Table 12.

**Table 12.** Confusion matrix for skin lesion classification using ResNet-50 (modified) CNN model.

|       | akiec | bcc   | bkl   | df    | mel   | nv    | vasc  |
|-------|-------|-------|-------|-------|-------|-------|-------|
| akiec | **97.90** | 0.24  | 0.97  | 0.63  | 0.03  | 0.07  | 0.16  |
| bcc   | 8.17  | **82.88** | 1.36  | 2.53  | 1.17  | 0.97  | 2.92  |
| bkl   | 7.82  | 0.45  | **87.33** | 3.05  | 0.27  | 0.27  | 0.81  |
| df    | 8.92  | 1.00  | 1.91  | **86.17** | 0.09  | 0.27  | 1.64  |
| mel   | 5.22  | 8.70  | 2.61  | 2.61  | **79.13** | 0.00  | 1.74  |
| nv    | 4.93  | 1.41  | 1.41  | 0.70  | 0.70  | **90.14** | 0.70  |
| vasc  | 5.50  | 3.06  | 2.45  | 3.67  | 0.92  | 0.31  | **84.10** |

As can be seen in Table 12, the highest classification accuracy was achieved for the actinic keratosis (akiec) skin pathology with accuracy equal to 97.90%, followed by nevus (nv) pathology with accuracy 90.14% and pigmented benign keratosis (bkl) skin pathology with the accuracy 87.33%. The skin pathologies which were found the most difficult among the evaluated ones to be correctly classified were melanoma (mel) and basal cell carcinoma (bcc) with classification accuracies equal to 79.13% and 82.88%, respectively. The experimental results indicate the potential of computer-aided diagnosis of skin lesions using image processing and machine learning technology. Based on the achieved classification accuracy, both in average (93.89%) and for the case of melanoma (79.13%), we deem that with the development of larger dermatoscopic image datasets and the progress in designing new

CNN architectures, new and more accurate classification models are feasible in support of the diagnosis of dermatologists.

## 5    Conclusion

An architecture for classification of skin lesions using convolutional neural networks was presented. The CNN-based architecture is based on pre-processing of dermatoscopic images followed by segmentation to extract the lesion area and subsequently processing of the corresponding segment of the image by a convolutional neural network for classification and labeling to a set of skin pathology types. In the evaluation new convolutional neural networks were trained using the Keras and TensorFlow python packets with CUDA supported. In addition, we modified and re-trained well known publicly available CNN models for image classification which have been pre-trained using big data collections. The experimental results showed that all evaluated modified and re-trained pre-existing CNN models outperformed the CNN models trained explicitly from the dermatoscopic image dataset used in the evaluation, with the best performing classification model being ResNet-50 with the accuracy equal to 93.89%. Analysis on skin lesion pathology type showed that melanoma and basal cell carcinoma were able to be correctly classified 79.13% and 82.88% of the times, respectively, while other types of pathologies like actinic keratosis and nevus were more easily correctly classified with accuracies 97.90% and 90.14%, respectively. The evaluation results show the potential of developing software tools for accurate computer-aided diagnosis of skin diseases which can be used as objective supportive tools for dermatologists.